\documentclass[aps,pra, onecolumn,14 pts,a4paper]{revtex4}
 \pdfoutput=1

\usepackage{amssymb,amsmath,amsfonts,graphicx}
\usepackage{bm}
\usepackage{epsfig,color,times}
 \usepackage{bm,color}
\usepackage{graphicx}
\usepackage{amssymb}
\usepackage{epstopdf}

\begin{document}

\title{ Equation for self-similar singularity of Euler 3D}

\author{Yves Pomeau} 
\affiliation{ Department of Mathematics, U. of Arizona,Tucson, AZ 85721, USA
}

\date{\today }

\begin{abstract}
The equations for a self-similar solution of an inviscid incompressible fluid are mapped into an integral equation which hopefully can be solved by iteration. It is argued that the exponent of the similarity are ruled by Kelvin's theorem of conservation of circulation. The end result is an iteration with a nonlinear term entering in a kernel given by a 3D integral (in general 3D flow) or 2D (for swirling flows), which seems to be within reach of present day computational power. Because of the slow decay of the similarity solution at large distances the total energy is diverging and recent mathematical results excluding a solution of the self-similar solution of Euler equation do not apply.

\end{abstract}

\maketitle

This note discuss the possibility of self-similar solution(s) of the Euler equations for an incompressible fluid in 3D. It follows a previous Note \cite{YP} on the case of axisymmetric flows and a discussion  \cite{modane} of the connection between singularities of the equations and recording in high speed wind tunnel.  
The set of equations we start from is 
\begin{equation}
{\partial_{t}{\bf{u}}} + {\bf{u}}\cdot \nabla {\bf{u}} = - \nabla p
\textrm{,}
\label{eq:Euler1}
\end{equation}
and 
\begin{equation}
\nabla \cdot {\bf{u}} = 0
\textrm{,}
\label{eq:Euler2}
\end{equation}
The vector  ${\bf{u}}$ is for the local value of the fluid velocity and $p$ is the pressure, a free function allowing to satisfy the condition of incompressiblity (\ref{eq:Euler2}). Moreover the nabla sign is for the gradient with respect to coordinate $ {\bf{r}}$. 

Years ago Leray wrote \cite{leray} the equations for a self-similar solution of the above set of equations, including by the way the viscosity. Our attempt is mostly limited to the inviscid (or Euler) case. It is believed, informally, that it is less likely to get a singularity with Navier-Stokes (namely with viscosity included) than with Euler, because of a possible regularizing effect of the viscosity. This is actually far more obvious for a small viscosity. A major question in the search of self-similar solution of the Euler equations is how to take into account the invariants of those equations. Actually there are two sets of invariants (excluding the total momentum which can be set to zero by convenient Galilean map. First we have conservation of energy $$  {\mathcal{E}} =  \frac{1}{2} \int \mathrm{d}{{\bf{r}}}  {\bf{u}}^2 \textrm{,}$$ 
This defines a quantity which is constant in the course of time if $ {\bf{u}}$ is a solution of the Euler equations. Then there are infinitely many conserved quantities which are the integrals $\int \mathrm{d} {\bf{s}} \cdot {\bf{u}} $ along any closed line of element $\mathrm{d} {\bf{s}}$, this line being carried by the flow. This Kelvin circulation theorem is a  highly non trivial property  and has important consequences in the search for self-similar solution of the Euler equation: if one assumes that such a closed line is carried by the flow inside the collapsing domain, Kelvin's theorem constrains the possible exponents of the similarity solution and, as we shall see, this constraint is not compatible (for point like singularities) with the conservation of energy inside the collapsing domain.  Let us consider axially symmetric flows and assume they remain so. Circles in planes perpendicular to the axis and with their center on the axis are mapped by the flow in circles with the same geometry. Therefore the circulation on such circles is conserved and so constrains the exponent of the similarity solution. 
 
 As did Leray \cite{leray} let us suppose there is a singularity in finite time by the evolution of this flow and that this singularity is of the self-similar type. It means that the corresponding solution of Euler equations is of the type $$ {\bf{u}}( {\bf{r}}, t) = (t^*- t)^{-\alpha}  {\bf{U}} ( {\bf{r}}(t^*- t)^{-\beta})\textrm{,}$$ where $t^*$ is the time of the singularity (taken as zero afterwards) and $\alpha$ and $\beta$ are positive exponents to be found and where  $ {\bf{U}}$ is to be derived by solving Euler equations. 
 
 That such a velocity field is a solution of Euler equation implies that $ 1 = \alpha + \beta$. The conservation of circulation implies that  $ 0 = \alpha - \beta$, and $\alpha = \beta = 1/2$. If one imposes instead that the energy is conserved in the collapsing domain, one must satisfy the constraint $- 2 \alpha + 3 \beta = 0$, which yields $\alpha = 3/5$ and  $\beta = 2/5$, the same exponents as for the Sedov-Taylor blast wave, although they correspond to a completely different situation. This shows that no set of singularity exponents can satisfy both constraints of energy conservation and of constant circulation on convected curves.  We shall choose, for a reason discussed later, the conservation of circulation. Notice that if the circulation is conserved the Reynolds number is also constant inside the collapsing domain, because it is the ratio of a typical value of the circulation (a constant) to the shear viscosity (another constant). This seems to go against one of the tenets of turbulence theory according to which once small scales are reached in the  cascade process the viscosity takes over and dissipates the energy. However with the scaling $\alpha =  \beta = 1/2$ the dissipated power in a singular event scales like  $ \nu  \int {\mathrm{d}} {\bf{r}} (\nabla {\bf{u}})^2  \sim (-t)^{-1/2}$ which is an integrable function of time showing that a finite amount of energy should be dissipated in each singularity event of the Navier-Stokes equation. 
 
 Let us introduce as new variable $ {\bf{R}} =  {\bf{r}}(- t)^{-1/2}$. The Euler equations become a set of equations for $ {\bf{U}} $ , a field function of  ${\bf{R}}$: 
 \begin{equation}
-\frac{1}{2}({\bf{U}}  + {\bf{R}}\cdot \nabla {\bf{U}}) + {\bf{U}}\cdot \nabla {\bf{U}} = - \nabla P
\textrm{,}
\label{eq:Euler1ss}
\end{equation}
\begin{equation}
\nabla \cdot {\bf{U}} = 0
\textrm{,}
\label{eq:Euler2ss}
\end{equation}
In both equations and later the nabla sign is for the gradient with respect to coordinate $ {\bf{R}}$.  

The behavior of ${\bf{U}}$ for $ {\bf{R}}$ large is derived from the constraint that, as $t$ tends to zero (namely the instant when the velocity field becomes singular at one point),  the solution of the similarity equation must become independent of $t$ at large distances, compared to the singular behavior of the diverging solution. This implies that ${\bf{U}}$ decays like $1/R $ in this limit times a vector function of the unit vector $\hat{\bf{R}} =  \frac{{\bf{R}}}{R}$. This yields an equation for the large distance behavior of the solution of equations  (\ref{eq:Euler1ss}) and (\ref{eq:Euler2ss}): because the velocity field decays like $1/R$ at large distances the nonlinear term, of order $R^{-3}$, becomes negligible in this limit. Let us put into equation (\ref{eq:Euler1ss}) a vector field $U({\bf{R}}) =  \frac{1}{R} {\bf{W}}_1 (\hat{\bf{R}})$. For such a field decaying like $1/R$ at large $R$ the linear term on the left-hand side of equation (\ref{eq:Euler1ss}) vanishes, as it should because it is the vanishing time derivative of the large distance behavior of the self-similar solution. However not any field of this type yields a convenient solution of the large distance behavior of the solution because it has also to satisfy the incompressibility condition (\ref{eq:Euler2ss}), a point to which we shall come back. 

Let us outline how the Laurent expansion of the solution of equations  (\ref{eq:Euler1ss}) and (\ref{eq:Euler2ss}) can be formally done. At leading order for $R$ large $U({\bf{R}}) \approx  \frac{1}{R} {\bf{W}}_1  (\hat{\bf{R}})$. Therefore it is natural (and this agrees with the algebra) to look for a solution with the following Laurent expansion: 
\begin{equation}
 {\bf{U}}({\bf{R}}) =  \sum_{n =1}^{\infty} \frac{1}{R^n} {\bf{W}_n}  (\hat{\bf{R}})
\textrm{.}
\label{eq:Euler2exp}
\end{equation}
Putting this expansion of ${\bf{U}}$ into equation  (\ref{eq:Euler1ss}), one finds: 
\begin{equation}
\frac{1}{2} \sum_{n =2}^{\infty}( n - 1)  \frac{1}{R^n} {\bf{W}_n}  (\hat{\bf{R}}) + {\bf{U}}\cdot \nabla {\bf{U}} = - \nabla P
\textrm{.}
\label{eq:Euler2exp.s}
\end{equation}
This equation is to be solved together with the incompressibility condition  (\ref{eq:Euler2ss}). This can be done, at least formally, by putting into the nonlinear term ${\bf{U}}\cdot \nabla {\bf{U}}$ the expansion given by equation (\ref{eq:Euler2exp}), including the term of order $1/R$ which is depends on a yet unknown function $W_1({\hat{R}})$. The writing of equation (\ref{eq:Euler2exp.s}) order by order with respect to $1/R$ yields two sets of equations. The first set of equations is derived by taking the gradient of equation (\ref{eq:Euler2exp.s}).  The algebra is made simpler by writing the other terms in this equation as infinite sums of inverse powers of $R$. This amounts to write: 
\begin{equation}
P({\bf{R}}) =  \sum_{k = 2}^{\infty} \frac{1}{R^k} P_k  (\hat{\bf{R}})
\textrm{.}
\label{eq:Euler2expP}
\end{equation}
and 
\begin{equation}
{\bf{U}}\cdot \nabla {\bf{U}} =  \sum_{m = 3}^{\infty} \frac{1}{R^m} {\bf{T}}_m (\hat{\bf{R}})
\textrm{.}
\label{eq:Euler2expRe}
\end{equation}
By identification of the same inverse power of $R$, one derives an equation for $P_k (\hat{\bf{R}})$ by taking the gradient of both sides of equation (\ref{eq:Euler2exp.s}) which yields: 
$$ - R \nabla^2 ( R^{-(k- 1)} P_{(k-1)} (\hat{\bf{R}})) =  \nabla\cdot  \frac{1}{R^k} {\bf{T}}_k(\hat{\bf{R}})\textrm{.}$$ Notice that, because of the condition of incompressiblity there is no contribution to this equation coming from the linear term on the left-hand side of equation (\ref{eq:Euler1ss}). This formula allows (in principle) to compute the components $P_k$'s  of the pressure by recursion. There remains to find a recursion formula for the components $ {\bf{W}}_n(\hat{\bf{R}})$ for $n> 1$. This is done by writing just equation (\ref{eq:Euler2exp.s}) as an equation for $ {\bf{W}}_n$ since the other functions of the same order, namely ${\bf{T}}_m (\hat{\bf{R}})$ are known in principle either by computing directly the nonlinear term or by solving Laplace's equation for the pressure at the right order with respect to $1/R$. The result is: 
$$\frac{1}{2}( n - 1) {\bf{W}}_n (\hat{\bf{R}}) +  {\bf{T}}_n(\hat{\bf{R}}) = - R^n  \nabla (R^{-(n-1)} P_{(n-1)} (\hat{\bf{R}}))\textrm{.}$$

This expansion in inverse powers of $R$ should be started with the solution of the linear problem, namely by knowing one way or another the function ${\bf{W}}_1 (\hat{\bf{R}})$. This function is constrained by the condition of incompressibility which reads: 
$$ \nabla \cdot ( \frac{1}{R}{\bf{W}}_1  (\hat{\bf{R}})) = 0 \textrm{.}$$ To give an example of function satisfying this constraint, take for instance, in Cartesian coordinates, $ \frac{1}{R} W_{1,x} =  \frac{y}{R^2}$ and $ \frac{1}{R} W_{1,y} =  - \frac{x}{R^2}$, where  $W_{1,x} $ is the $x$-component of the vector ${\bf{W}}_1$. This vector field is divergence-free.  More generally such a divergence free vector field is, in the present case (namely for a vector of modulus depending on the modulus of $R$ like $1/R$) a vector field that is the curl of an arbitrary smooth vector field $\bf{A} ({\hat{R}})$.  

The large distance behavior of ${\bf{U}}$ as $1/R$ puts this kind of solution of the similarity equation outside of the range of applicability of published mathematical results on the possibility of a solution of this equation. According to Chae \cite{Chae} no solution exists if the vorticity belongs to a function space $L^p$ with $p$ close to $0$. In the present case the vorticity decays like $R^{-2}$ at $R$ large so that the vorticity field does not belong to $L^p$ for $p$ close to zero. Therefore this kind of result does not be apply to the kind of solution we consider.

Even though this note is mostly concerned with the existence of solutions of the Euler inviscid equation, let us consider the existence  of self-similar singular solution of the Navier-Stokes  equation with viscosity, the problem dealt with by Leray. In this case the equation (\ref{eq:Euler1ss}) is replaced by an equation with the viscosity included: 
 \begin{equation}
-\frac{1}{2}({\bf{U}}  + {\bf{R}}\cdot \nabla {\bf{U}}) + {\bf{U}}\cdot \nabla {\bf{U}} = - \nabla P +  \nu  \nabla^2{\bf{U}} 
\textrm{,}
\label{eq:Euler1ssL}
\end{equation}
This was the equation originally given by Leray, $\nu$ being the shear viscosity of the fluid. Because the viscosity term involves second derivatives it is subdominant compared to the streaming term  $\frac{1}{2}({\bf{U}}  + {\bf{R}}\cdot \nabla {\bf{U}}) $ on the left-hand side for solutions decaying algebraically at large $R$. Therefore one can use the result derived above for the asymptotic behavior of solutions of  Leray equation at $R$ large, which yields ${\bf{U}} \sim 1/R$ at large $R$. This estimate is interesting because it puts the velocity field outside of the $L^3$ function space since the corresponding integral diverges logarithmically at large $R$. The non existence of a solution of Leray equation in 3D has been proved \cite{serbes} under the condition that the velocity field belongs to the space $L^3$, a condition seemingly incompatible with the asymptotic behavior derived from the equations themselves.  

Let us outline a way of approaching the numerical search of a solution of equations (\ref{eq:Euler1ss}) and (\ref{eq:Euler2ss}). Taking the divergence of equation (\ref{eq:Euler1ss}) one finds the Reynolds equation relating the pressure to the velocity field: 
 \begin{equation}
\nabla^2 P + \partial_i (U_j \partial_j U_i) = \nabla^2 P + (\partial_i U_j) (\partial_j U_i) = 0
\textrm{,}
\label{eq:Euler3ss}
\end{equation}
an expression where we used the symbol $\partial_i  = \frac{\partial}{\partial X_i}$, the indices $i$ and $j$ being for the Cartesian coordinates with the convention of summation over repeated indices. In an unbounded geometry like the present one, the Poisson equation for $P$ can be solved like: 
 \begin{equation}
P({\bf{X}}) = \frac{1}{4 \pi}   \int  {\mathrm{d}} {\bf{X}}'   \frac{1}{\vert  {\bf{X}}'  -  {\bf{X}} \vert } (\partial_i U_j') (\partial_j U_i')
\textrm{,}
\label{eq:Euler4ss}
\end{equation}
In the expression above $U'_i$ is for the $i$-component of the vector ${\bf{U}}({\bf{X}}')$.  The integral in equation  (\ref{eq:Euler4ss}) converges at large $({\bf{X}}')$ because ${\bf{U}}$ decays like $1/R$ at large distances.  From this solution of Poisson equation one derives the gradient of the pressure: 
 \begin{equation}
 \partial_k  P = \frac{1}{4 \pi}   \int  {\mathrm{d}} {\bf{X}}'   \frac{X_k'  -  X_k}{\vert  {\bf{X}}'  -  {\bf{X}} \vert ^3}  (\partial_i U_j') (\partial_j U_i') \textrm{,}
\label{eq:EulerP}
\end{equation}
 From this one may derive an integral equation for a solution  ${\bf{U}}({\bf{X}})$ of the similarity equations. Let us introduce the vector ${\bf{V}}({\bf{X}})$ with Cartesian components: 
 \begin{equation}
V_k({\bf{X}}) =  \partial_k  P +  U_j \partial_j U_k
\textrm{,}
\label{eq:EulerV}
\end{equation}
Putting there the pressure gradient given by equation (\ref{eq:EulerP}) on finds ${\bf{V}}$ as a quadratic functional of  ${\bf{U}}$. 
Once $V_k({\bf{X}})$ is given it is easy to compute the velocity field ${\bf{U}}$ by solving formally equation (\ref{eq:Euler1ss}) in spherical coordinates. This yields: 
 \begin{equation}
U_k({\bf{X}}) = \frac{1}{X}   \int_0^{X}  {\mathrm{d}} X' (- 2 V_k(X', {\hat{\bf{X}}}))
\textrm{,}
\label{eq:EulerUsol}
\end{equation}
In the expression above ${\hat{\bf{X}}} = \frac{{\bf{X}}}{X}$, $X$ being the length of ${\bf{X}}$. This gives the idea of a solution of equation  (\ref{eq:Euler1ss}) by iteration, namely by putting on the right-hand side of equation (\ref{eq:EulerUsol}) a trial function ${\bf{U}}$ satisfying the same known constraints as the solution we are looking for (incompressibility and asymptotic behavior), then computing this right-hand-side and taking it as initial condition for another iteration, etc. Of course there is no guarantee that this iteration will converge to a fixed point but if it does one would have found a solution of the similarity equation. Such a solution, if it exists, is  not unique because the set of solutions of (\ref{eq:Euler1ss}) has a continuous symmetry: if  ${\bf{U}}({\bf{X}})$ is a solution, then $ \lambda{\bf{U}}( \lambda{\bf{X}})$, $ \lambda$ arbitrary real constant is also a solution, the pressure being multiplied by  $\lambda^2$. Therefore in a numerical search of the solution(s?) by iteration of equation (\ref{eq:EulerUsol}) one has to get rid of a possible idling between different and more or less random values of $ \lambda$. To avoid this one should normalize, for instance to one, the solution at each step of the iteration. This normalization cannot be done with the $L^2$ norm of ${\bf{U}}({\bf{X}})$. The energy of the solution diverges at $R$ large because this solution decays like $1/R$ at $R$ large. This explains why this self-similar solution does not have to conserve the energy, just because its energy diverges and is so undefined. A possible way of normalizing the solution is to take its $L^4$ norm and make it constant from one step of the iteration to the next,  for instance equal to $1$.  Because the problem is formally invariant by rotation, there is another possibility of idling of solutions by random rotation from one step of the iteration to the next. How to avoid this random rotation is more difficult and technical. We shall deal with this issue in a coming publication. 

This symmetry allows to find  singular solutions of Euler equation of finite energy. Because the velocity decays like $1/R$ at $R$ large, the energy of a self-similar solution diverging at only one point diverges. This divergence can be eliminated by considering a solution made of the addition of two point-wise solutions diverging at very distant points and at the same time. The solutions at each point  are not the same, but mapped into each other by the dilation transform with $\lambda = -1$. Thanks to that the two solutions at distances far bigger than the distance between the two points of singularity cancel at leading order, so that the net result is a velocity field decaying at least like $1/R^2$ far from the two points of singularity, which makes the energy convergent.

It follows from equation (\ref{eq:EulerUsol}) that the vector called ${\bf{W}}_1({\hat{\bf{R}}}) $ before, which is the coefficient of $1/R$ in the leading order term of the expansion of ${\bf{U}}$ at $R$ large, is related to an integral of ${\bf{V}}$ as: 
$$ {\bf{W}}_1 ({\hat{\bf{R}}}) =  \int_0^{\infty}  {\mathrm{d}} X' (- 2 {\bf{V}}(X', {\hat{\bf{R}}})) \textrm{,}$$

It is also of interest to notice that the equations are obviously consistent with an axisymmetric geometry. In this case the number of components of ${\bf{R}}$ is reduced to two.  

Among other things, the success of an iterative solution is strongly linked to the dimensionality of space where the integral on the right-hand side of  equation (\ref{eq:EulerUsol}) has to be performed: the bigger is this dimension likely the more difficult it is to find a fixed point by iteration. This depends on which contribution to ${\bf{V}}$ one considers. The contribution  $U_j \partial_j U_k$ is just integrated once over the length of ${\bf{X}}$, which yields a one dimensional integral. Given the velocity field  ${\bf{U}}$, the contribution of the pressure gradient requires a priori integration on four dimensions: three dimensional integration as resulting from equation (\ref{eq:EulerP}) to obtain the pressure gradient from the velocity field, plus the other one dimensional integration  in equation (\ref{eq:EulerUsol}). Notice however that an axisymmetric swirling flow suppress one dimension for the space variable and so yield an integral operator with one dimension less than in the general case. 

It is perhaps of interest to notice that the integration on the length of ${\bf{X}}$ can be done analytically because of the way the coordinate ${\bf{X}}$ enters into  equation (\ref{eq:EulerP}). This integral requires to know the value of the following  definite integrals: 
$$ I_{1, k} ({\bf{X}}', {\hat{\bf{X}}}) =  \int_0^{X}  {\mathrm{d}} X''  \frac{X'' {\hat{\bf{X}}}_k}{\vert  X'' {\hat{\bf{X}}}  -  {\bf{X}}' \vert ^3}  \textrm{,}$$
and 
$$ I_{2} ({\bf{X}}', {\hat{\bf{X}}}) =  \int_0^{X}  {\mathrm{d}} X''  \frac{1}{\vert  X'' {\hat{\bf{X}}}  -  {\bf{X}}' \vert ^3}  \textrm{.}$$
The result of both integrations can be expressed by means of incomplete elliptic integrals and their derivatives.  However it is not obvious that much is gained from such a formal integration. It could be better to eliminate the formal singularity of the integral in equation (\ref{eq:EulerP}) for $ {\bf{X}}'  =  {\bf{X}}$ by changing integration variable from ${\bf{X}}' $ to ${\bf{X}}''  = {\bf{X}}' - {\bf{X}}$ and integrating in spherical coordinates for ${\bf{X}}''$ in order to eliminate formally any zero denominator. This would yield :  
 \begin{equation}
 \partial_k  P = \frac{1}{4 \pi}   \int  {\mathrm{d}} {\bf{X}}''  \frac{X_k'''}{\vert  {\bf{X}}'' \vert ^3}  (\partial_i U_j') (\partial_j U_i') ( {\bf{X}}''  +  {\bf{X}}')  \textrm{,}
\label{eq:EulerPch}
\end{equation}
By writing in spherical coordinates $$  {\mathrm{d}} {\bf{X}}''  \frac{X_k''}{\vert  {\bf{X}}'' \vert ^3}  =  {\mathrm{d}}X ''   \  {\mathrm{d}}  {\hat{\bf{X}}}''   \ {\hat{\bf{X}}}_k'' \textrm{,} $$ one sees that the vanishing denominator is eliminated. 

An iterative solution of equation (\ref{eq:EulerUsol})  is surely a non trivial endeavor. From the point of view of a mathematical approach of this question, it should be observed that the number of derivatives and of integrations on the right-hand side of this equation is the same. Therefore, qualitatively the functions obtained after each iteration step have no obvious tendency to have more and more small scale oscillations, as it would perhaps be the case if the order of derivation increases at each step of the iteration.  

{\underline{Acknowledgments}}

This work has benefited of discussions with Jean Ginibre, Martine Le Berre, B\'ereng\`ere Dubrulle, Christophe Josserand and Thierry Lehner.  
  
{}
\end{document}